\documentclass[fleqn,10pt]{wlscirep}
\usepackage[utf8]{inputenc}
\usepackage[T1]{fontenc}
\usepackage{bm}
\usepackage{graphicx}
\usepackage{subfigure}
\usepackage[euler]{textgreek}
\usepackage{ulem}
\usepackage{float}
\usepackage{color}
\usepackage[version=3]{mhchem}
\usepackage{verbatim}

\title{Raman scattering from the bulk inactive out--of--plane B$^{1}_{2\text{g}}$ mode in few--layer MoTe$_{2}$}

\author[1,*]{M.~Grzeszczyk}
\author[1]{K.~Go\l{}asa}
\author[1,2]{M.~R.~Molas}
\author[1,2]{K.~Nogajewski}
\author[1]{M.~Zinkiewicz}
\author[1,2]{M.~Potemski}
\author[1]{A.~Wysmo\l{}ek}
\author[1]{A.~Babi\'nski}

\affil[1]{Faculty of Physics, University of Warsaw, ul. Pasteura 5, 02-093 Warsaw, Poland}
\affil[2]{Laboratoire National des Champs Magn\'etiques Intenses, CNRS-UGA-UPS-INSA, 25, Avenue des Martyrs, 38042 Grenoble, France}

\affil[*]{magdalena.grzeszczyk@fuw.edu.pl}

\begin{abstract}

Raman scattering from the out-of-plane vibrational modes (A$_{1\text{g}}$/A'$_{1}$), which originate from the bulk-inactive out--of--plane B$^{1}_{2\text{g}}$ mode, are studied in few-layer \ce{MoTe2}.
Temperature-dependent measurements reveal a doublet structure of the corresponding peaks in the Raman scattering spectra of tetralayer and pentalayer samples. A strong enhancement of their lower energy components is recorded at low temperature for 1.91~eV and 1.96~eV laser excitation. We discuss the attribution of the peaks to the inner modes of the respective Raman-active vibrations. The temperature evolution of their intensity strongly suggests a resonant character of the employed excitation, which leads to the mode enhancement at low temperature. The resonance of the laser light with the singularity of the electronic density of states at the \textit{M} point of the Brillouin zone in \ce{MoTe2} is proposed to be responsible for the observed effects.
\end{abstract}

\begin{document}

\flushbottom
\maketitle

\thispagestyle{empty}

\section*{Introduction}


Transition metal dichalcogenides (TMDs) are layered materials, which gather growing interest of research community. This is related to their potential applications in electronics \cite{binder2017sub,roy2014field}, optoelectronics \cite{mak2016photonics}, photovoltaics \cite{bernardi2013extraordinary, wong2017high} and thermoelectricity \cite{yoshida2016gate}, just to mention a few topics of the utmost importance to the solid-state society. A specific structure of TMDs comprising layers of covalently bound metal and chalcogen atoms, which are kept together by relatively weak van der Waals forces makes these materials very interesting also from fundamental point of view. It is reflected in particular in their lattice dynamics, which can be revealed by Raman scattering (RS) spectroscopy \cite{saito}.  Both low-energy shear and breathing modes \cite{zhao12,zhang11} and high-energy vibrational modes \cite{lee, li2012bulk, golasa2014resonant,chakraborty2013layer} critically depend on the number of layers, which form the sample under study. The effect of the sample's thickness on the RS can be even stronger when resonant excitation is employed. The resulting change in the electronic configuration of the investigated material leads to several effects, which are not usually observed in the non-resonant RS spectroscopy such as the enhancement of particular features in the RS spectrum \cite{livneh16} related to multiphonon processes \cite{golasa} or quenching of Raman-active modes due to quantum interference \cite{golasa2017,miranda2017quantum}. The observation of Davydov splitting, which arises from interlayer interactions between two equivalent chalcogen-metal-chalcogen layers forming the unit cell of 2H--TMDs and vibrating with a non-zero phase shift one with respect to another, is also facilitated by resonant excitation of the RS spectra \cite{davydov1964theory,wermuth1979glossary}. It has been reported for all semiconducting few-layer TMD materials: \ce{MoS2} \cite{molina2015vibrational}, \ce{MoSe2} \cite{tonndorf}, \ce{WS2} \cite{staiger}, \ce{WSe2} \cite{kim2017excitonic} and \ce{MoTe2} \cite{froehlicher,grzeszczyk,song2016}. Resonant RS also unveils otherwise inactive vibrational modes of the crystal lattice. An example of such "brightenining" is the observation of the out-of-plane modes, which originate from the bulk-inactive B$^{1}_{2\text{g}}$ vibrations in TMDs \cite{yamamoto, guo, terrones2014new}. 

In this work we investigate the RS in few-layer \ce{MoTe2} by employing various excitation wavelengths. Temperature evolution of the RS spectra in 2- to 5-layer-thick \ce{MoTe2} is studied for 1.91~eV and 1.96~eV laser light excitation. A doublet structure of the out-of-plane mode, which can be hardly noticed at room temperature, becomes clearly observable at low temperature in tetra- and pentalayer samples. This results from a substantial increase of its lower-energy component's intensity with decreasing temperature. The experimental results are discussed with respect to a previously reported temperature evolution of the A$_{1\text{g}}$/A'$_{1}$ mode in \ce{MoTe2} \cite{golasa2017}. We conclude that the observed temperature evolution of the investigated out-of-plane mode is related to resonant character of the RS for the employed light excitation. We propose that the resonance of the laser light with a singularity of the electronic density of states at the \textit{M} point of the Brilloiun zone in \ce{MoTe2} might be responsible for the observed anomalous increase of the RS features.

\section*{Results}


\begin{figure}
	\centering  
		{\includegraphics[width=0.48\linewidth]{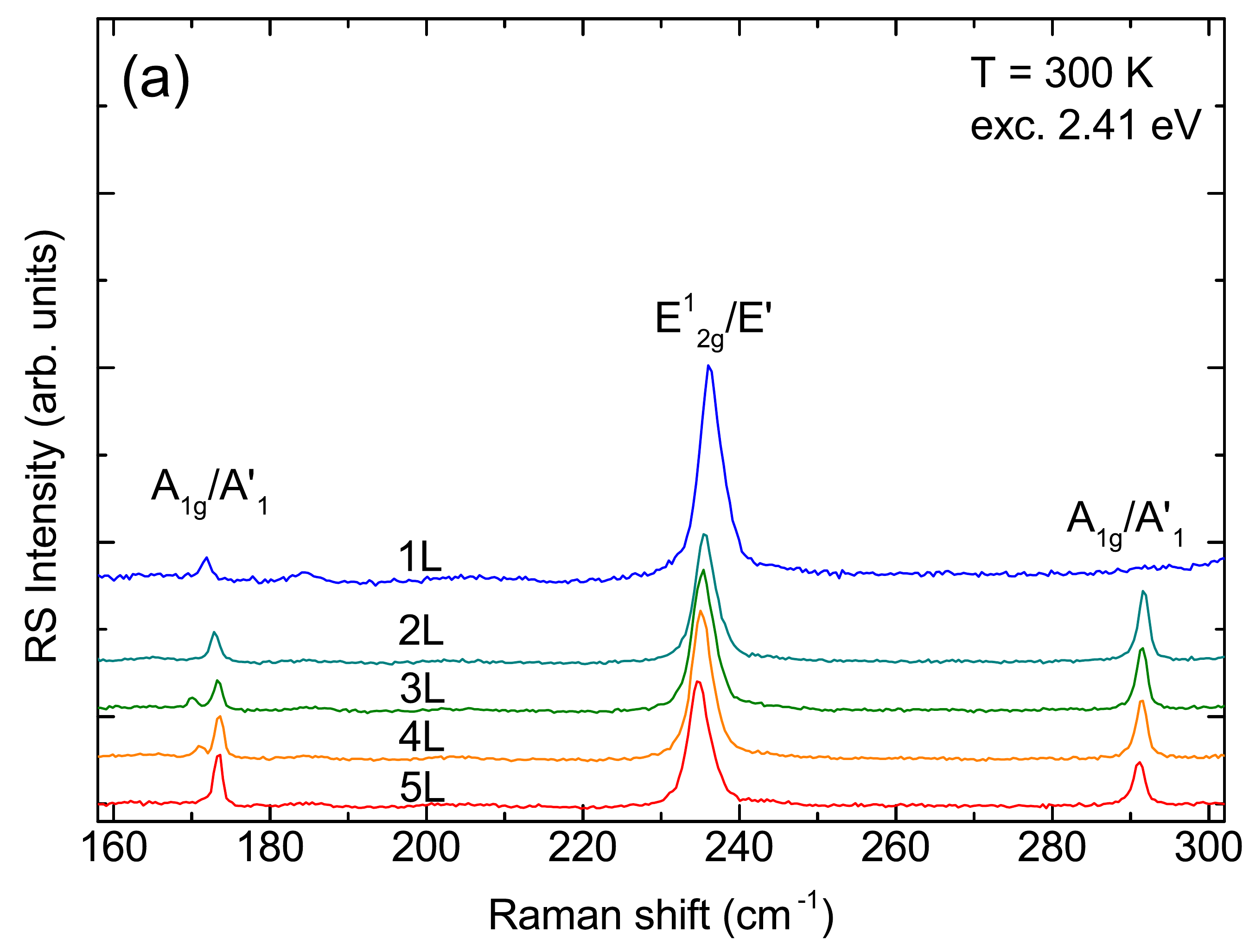}}
		{\includegraphics[width=0.48\linewidth]{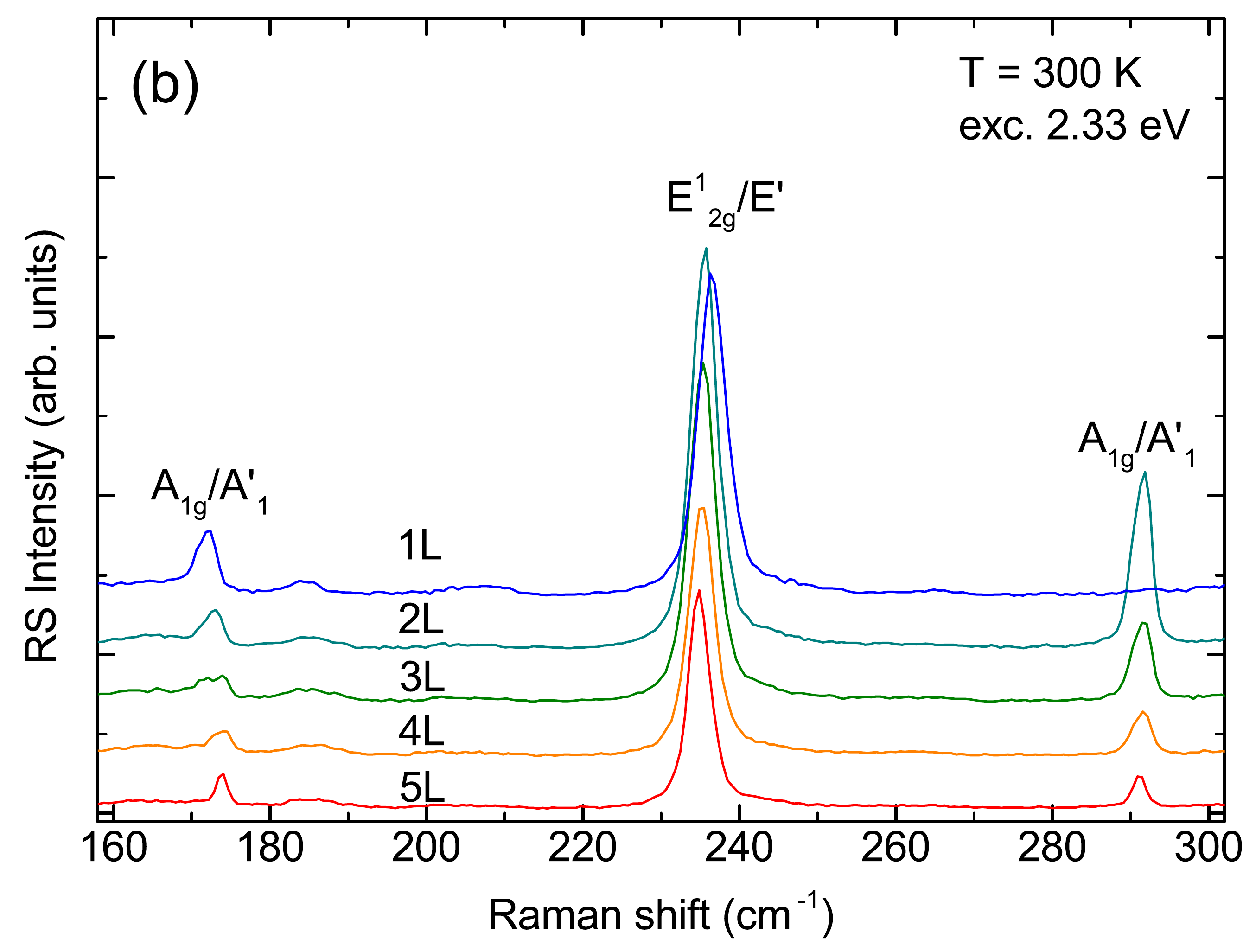}}
		{\includegraphics[width=0.48\linewidth]{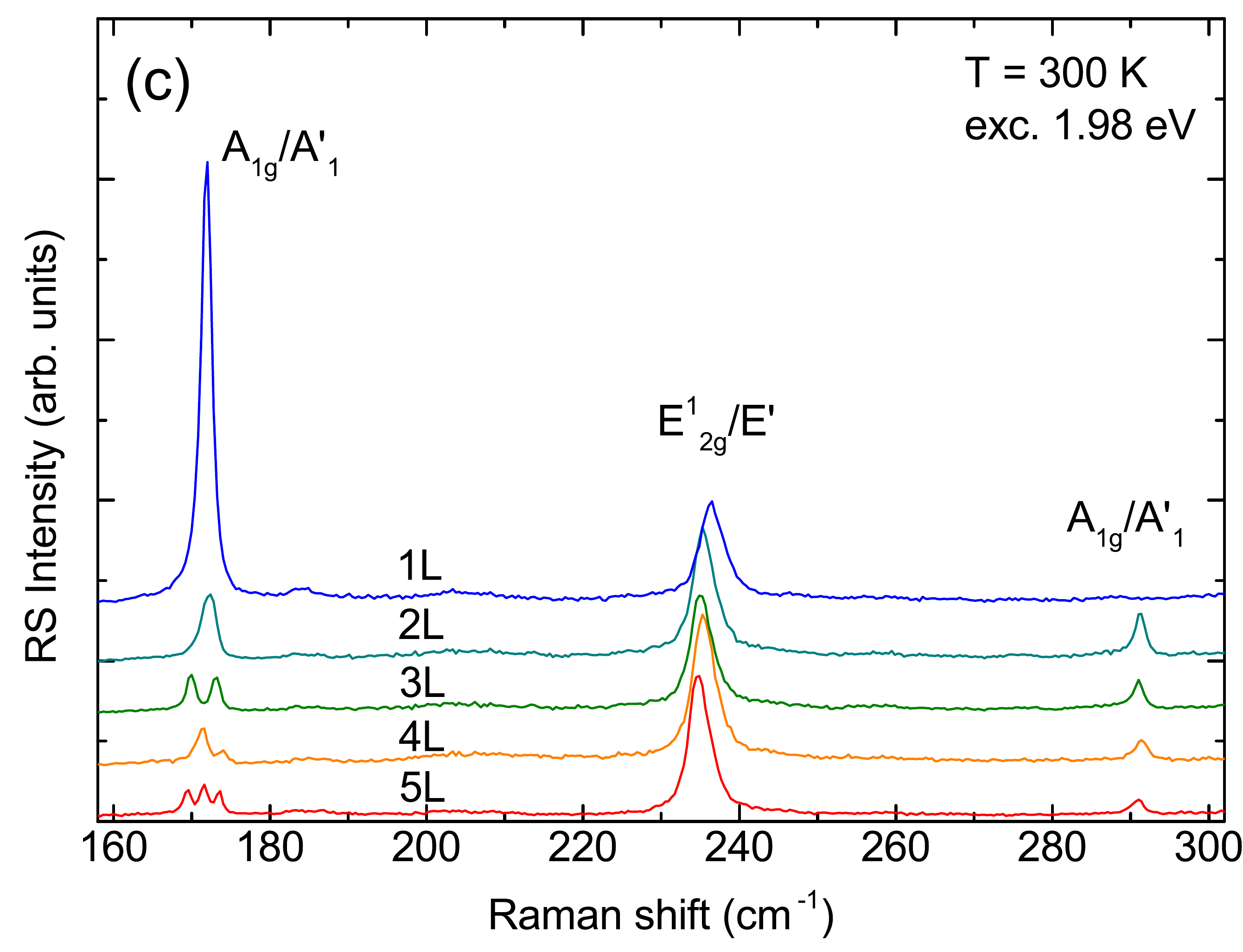}}
		{\includegraphics[width=0.48\linewidth]{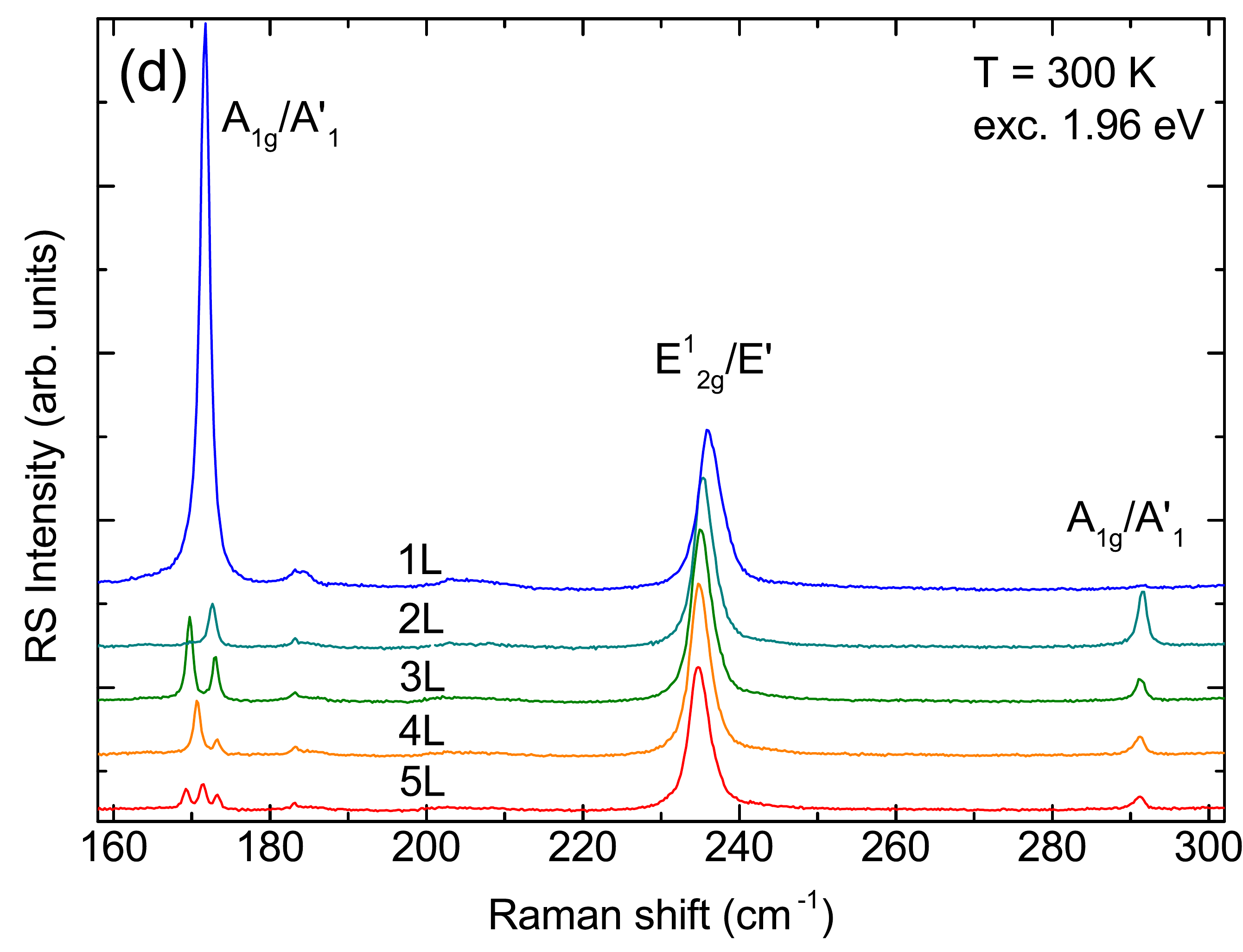}}
		{\includegraphics[width=0.48\linewidth]{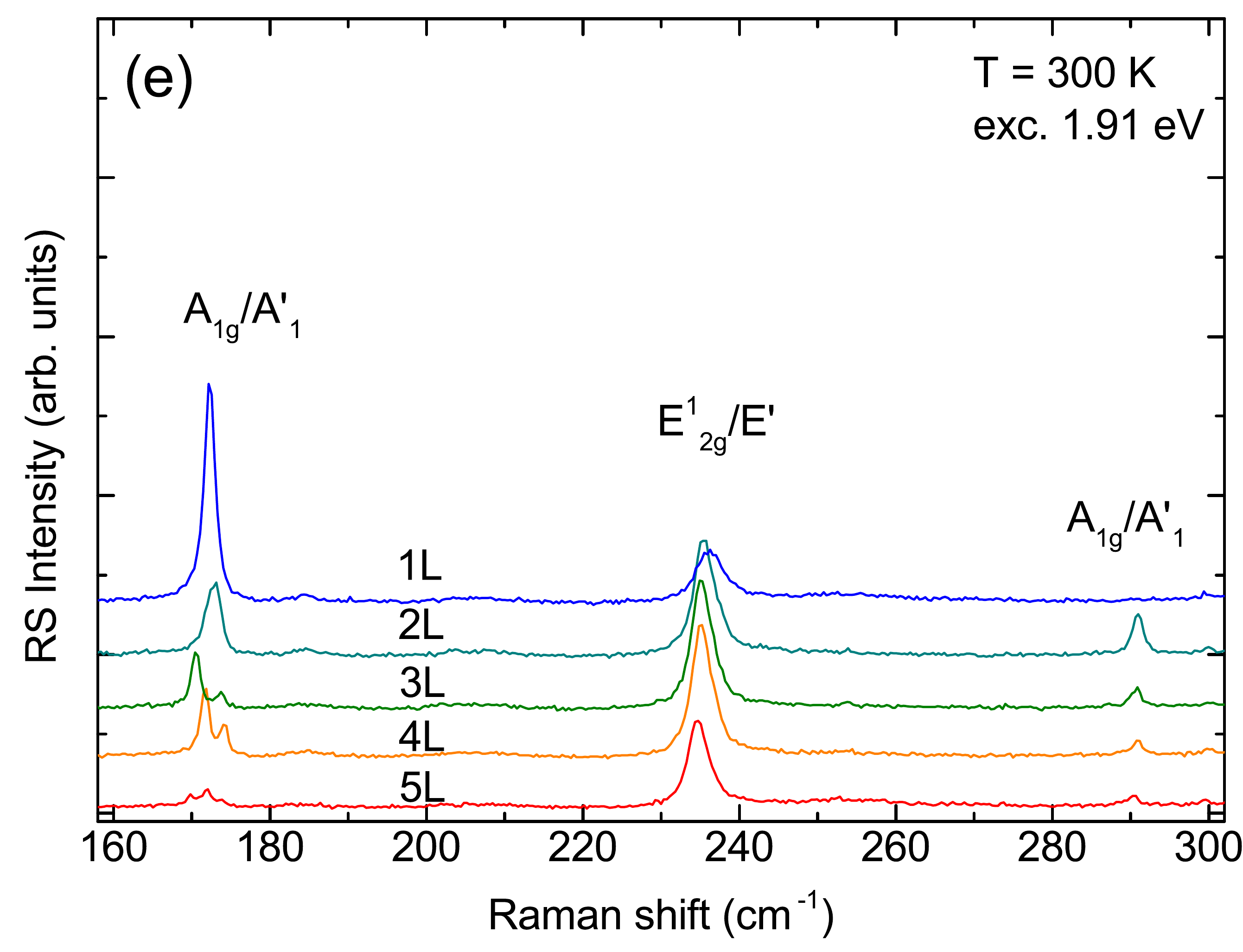}}
		{\includegraphics[width=0.48\linewidth]{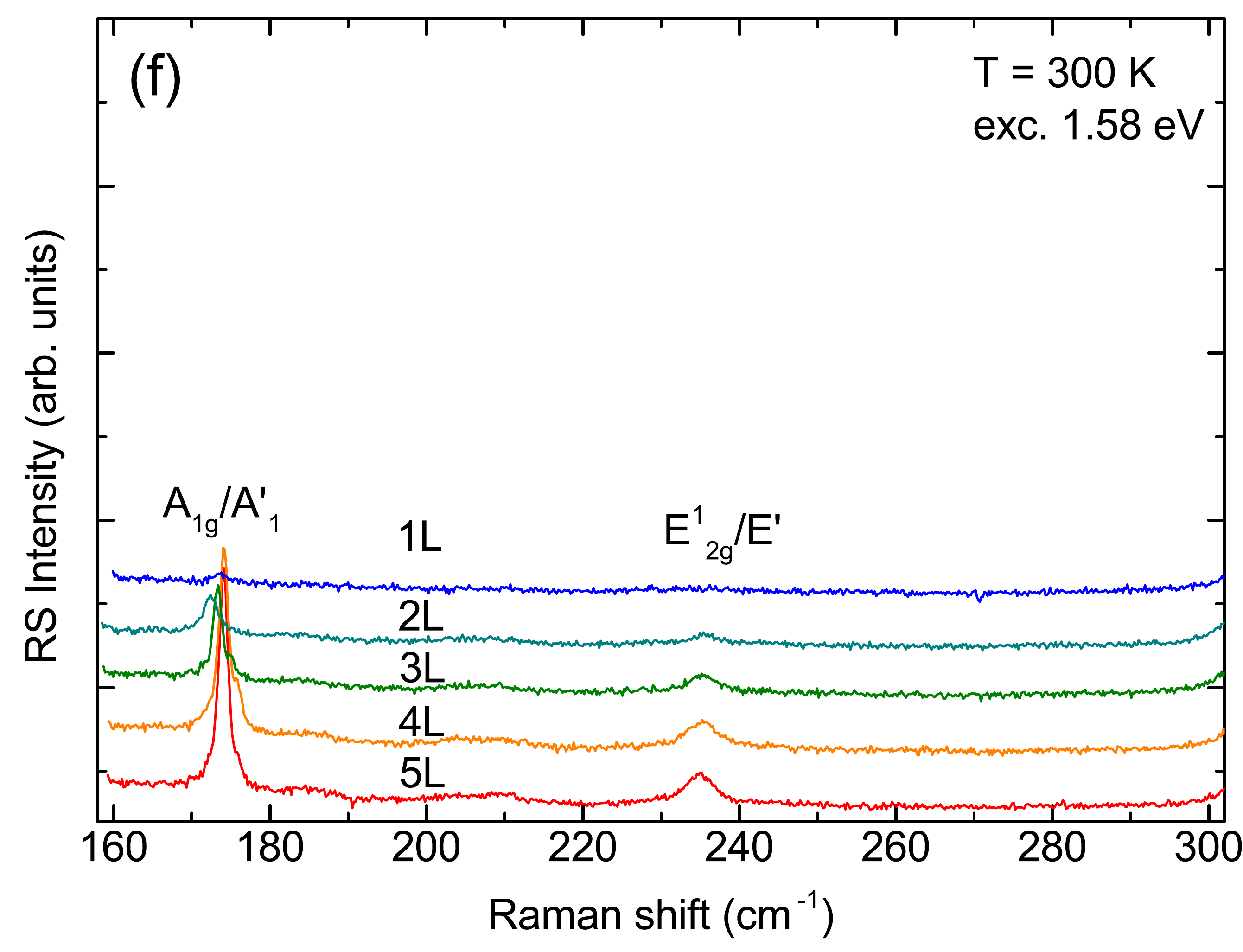}}
	\caption{Raman scattering (RS) spectra of monolayer (1~L) to pentalayer (5~L) \ce{MoTe2} measured at room temperature using various excitation energies.}
	\label{fig:collection}
\end{figure}

Molybdenum telluride (\ce{MoTe2}) is a semiconductor TMD with a relatively small band-gap energy close to 1.3~eV \cite{lezama2015indirect} and with a spin--orbit coupling, which is stronger than in other molybdenum-based TMDs \cite{pradhan2014field}. These properties make it interesting for a multitude of potential applications \cite{late2012,jariwala,val,mak2014}. Moreover, the small energy gap of \ce{MoTe2} opens up a possibility to study its resonant RS with fairly standard light sources like He-Ne (1.96~eV) or Kr-Ar (1.91~eV) lasers. The former laser energy coincides, in particular, with substantial density of states in few-layer \ce{MoTe2}, which occurs at the \textit{M} and \textit{K} points of the Brillouin zone \cite{guo}. Similar to other semiconducting tungsten- and molybdenum-based TMDs, \ce{MoTe2} crystallizes in a trigonal prismatic (2H) structure, whose symmetries in the limit of bulk material belong to the D$_{6h}$ point group. As a result, in multilayer \ce{MoTe2} there are four Raman-active modes: in-plane E$_{1\text{g}}$, E$^{1}_{1\text{g}}$, E$^{2}_{2\text{g}}$, and out-of-plane A$_{1\text{g}}$, as well as four IR-active modes: E$^{1}_{1\text{u}}$, E$^{2}_{1\text{u}}$, A$^{1}_{2\text{u}}$, and A$^{2}_{2\text{u}}$.
Other four modes E$_{2\text{u}}$, B$_{1\text{g}}$, B$^{1}_{2\text{g}}$, B$_{1\text{u}}$ are optically inactive. Point groups describing symmetries of few-layer \ce{MoTe2} composed of even and odd number of layers are D$_{3d}$ and D$_{3h}$, respectively. This results in different symmetry labelling of vibrational modes. In a structure with odd (even) number of layers the Raman-active modes are denoted by E', E'', and A'$_{1}$ (E$_{\text{g}}$ and A$_{1\text{g}}$). It refers to modes observed in the RS spectra of few-layer \ce{MoTe2} as presented in Fig. \ref{fig:collection}, which demonstrate the results obrained for monolayer (1~L) to pentalayer (5~L) \ce{MoTe2} excited with several different laser wavelenghts. It can be seen that both the relative intensities and spectral lineshapes of individual peaks for a particular thickness strongly depend on the excitation energy. 

Let us focus first on the peaks due to the out-of-plane A$_{1\text{g}}$/A'$_{1}$ vibrations at $\sim$174~cm$^{-1}$, which correspond to the Raman--active A$_{1\text{g}}$ mode in bulk \ce{MoTe2} \cite{froehlicher,grzeszczyk,song2016}. The Davydov splitting of the mode can be clearly observed in 3~L, 4~L and 5~L \ce{MoTe2} for 1.98~eV, 1.96~eV or 1.91~eV, while it remains hardly distinguishable for 1.58~eV excitation. Relative intensities of the lower- and higher-energy components of the A'$_{1}$ mode in 3~L \ce{MoTe2} also depend on the excitation energy. The higher-energy component of the A'$_{1}$ mode in 3~L \ce{MoTe2}, which is due to vibrations occuring with the same phase in all three covalently bound \ce{MoTe2} layers is of higher intensity than its lower-energy counterpart in the RS spectrum excited with 2.41~eV light. On the contrary, the lower--energy component due to the mode in which the tellurium atoms in the middle \ce{MoTe2} layer vibrate out-of-phase as compared to vibrations in the outer layers, is of higher intensity in the RS spectrum excited with 1.96~eV or 1.91~eV light. Finally, the Davydov--split components of the A'$_{1}$ mode in 3~L \ce{MoTe2} are of similar intensity in the RS spectrum excited with 1.98~eV light. As compared to the in-plane mode, the relative intensities of spectral features originating from the out-of-plane mode also strongly depend on the excitation energy. For 1~L \ce{MoTe2} the out-of-plane mode dominates the RS spectrum excited with 1.98~eV, 1.96~eV or 1.91~eV light, whereas the A$_{1\text{g}}$/A'$_{1}$ peak is much weaker than the one due to in-plane vibrations corresponding to the E$^{1}_{2\text{g}}$ mode in the bulk (observed at $\sim$235 cm$^{-1}$). For 5~L \ce{MoTe2} the out-of-plane mode is weaker than its in-plane counterpart for 2.41~eV and 2.33~eV excitation while the opposite takes place when 1.58~eV excitation is employed. This behaviour points at the resonant RS of light on out-of-plane modes, which occurs when the laser energy coincides with a maximum of electronic density of states in the crystal lattice \cite{guo}. This effect was previously reported for \ce{MoTe2} \cite{song2016,grzeszczyk,golasa2017anomalous} and explained in terms of quantum interference of contrubutions from different points of the Brillouin zone and the electron-phonon coupling \cite{miranda2017quantum}. A possible role of resonant vs non-resonant contributions to the RS efficency was also suggested in Ref. \citenum{golasa2017}.

\begin{figure}[h]
	\centering
		\includegraphics[width=\textwidth]{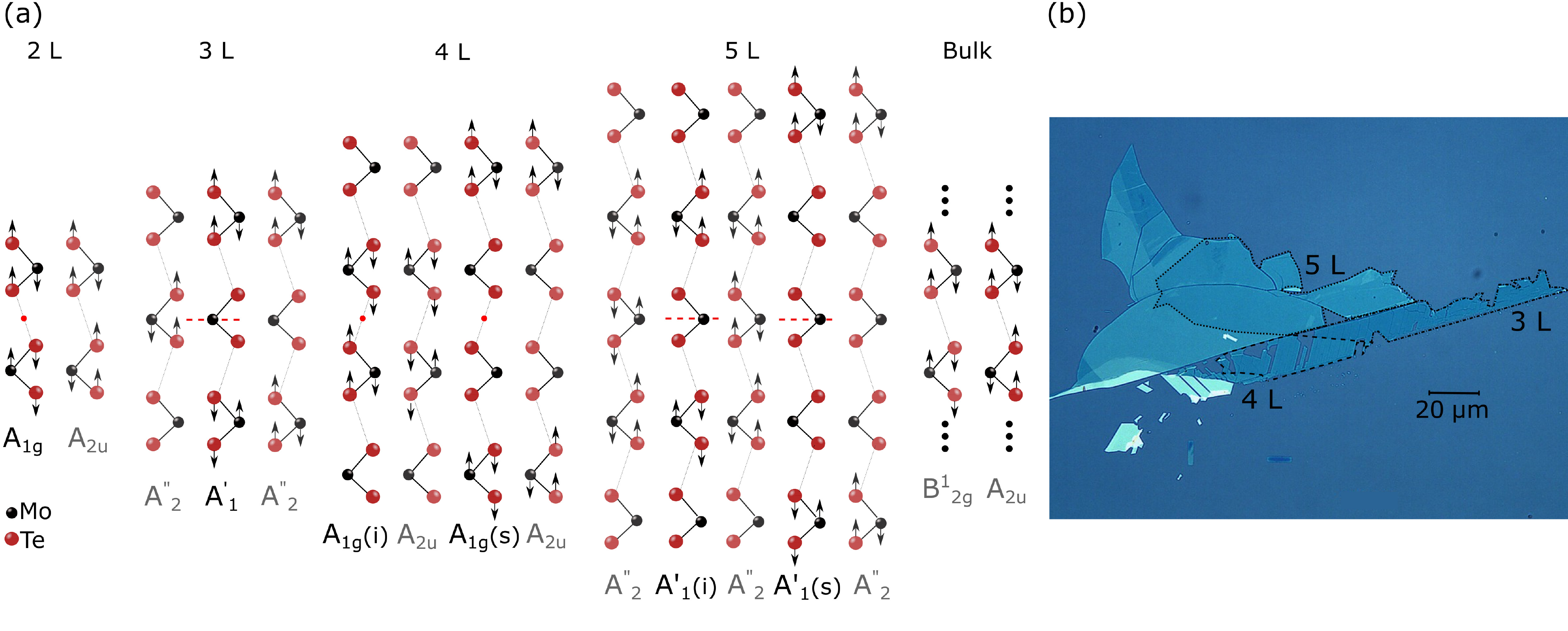}
	\caption{(a) Schematic representation of vibrational modes in 2 - 5~L \ce{MoTe2}. The corresponding bulk-inactive B$^{1}_{2\text{g}}$ mode is also shown for comparison. Dashed red lines denote the mirror symmetry plane and the red points mark the inversion centers. (b) Microscopic image of the studied \ce{MoTe2} flakes.}
\label{fig:mody}
\end{figure}

\begin{figure}
	\centering  
		\includegraphics[width=0.8\linewidth]{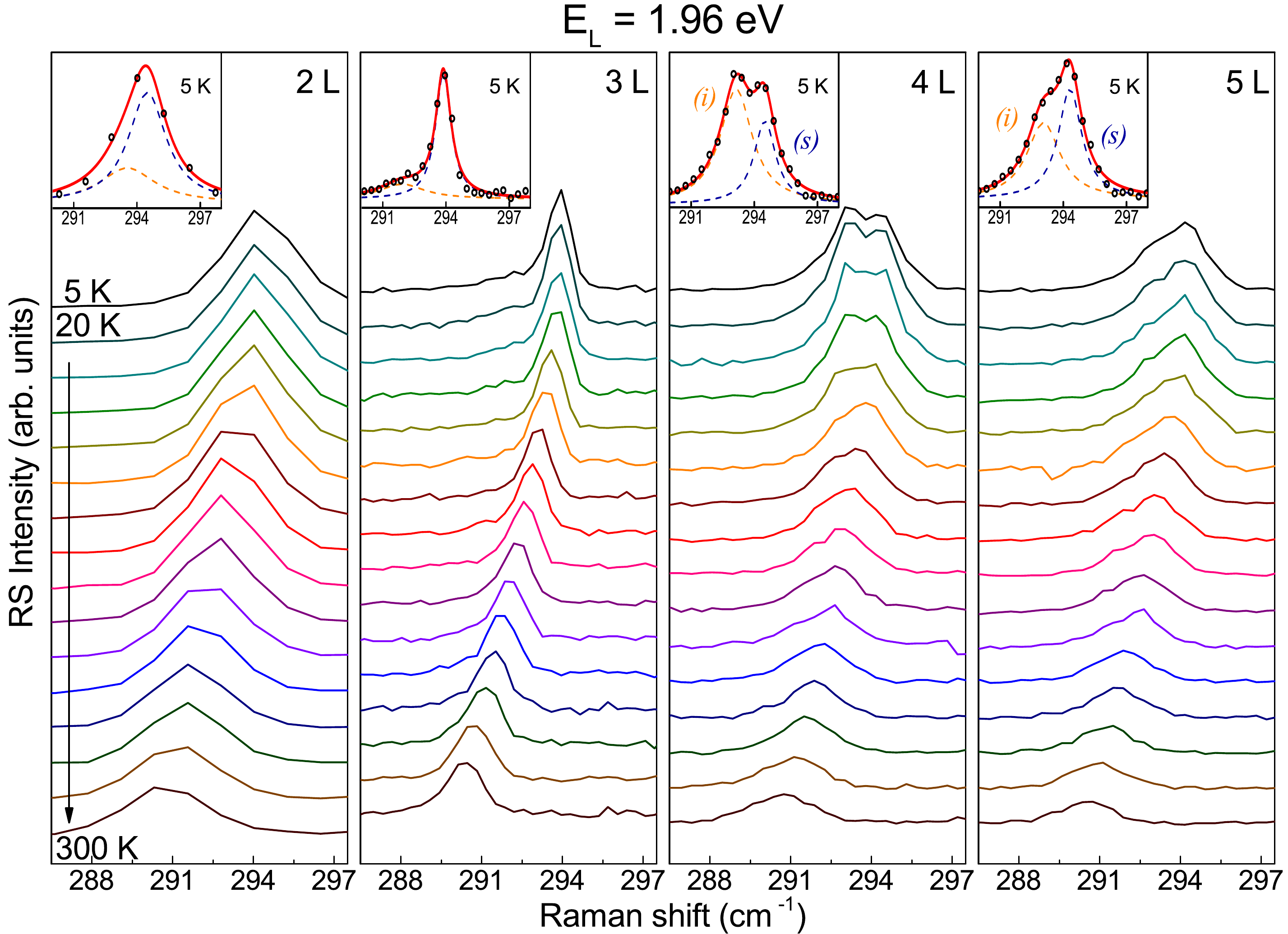}
		\caption{Temperature evolution of high-energy part of the RS spectra of  2 -- 5~L \ce{MoTe2} with the out-of-plane A$_{1\text{g}}$/A'$_{1}$ modes corresponding to the bulk inactive B$^{1}_{2g}$ mode. Laser excitation energy equals 1.96~eV ($\lambda$~=~632.8~nm).}
	\label{fig:633temp}
\end{figure}

\begin{figure}
	\centering  
		\includegraphics[width=0.8\linewidth]{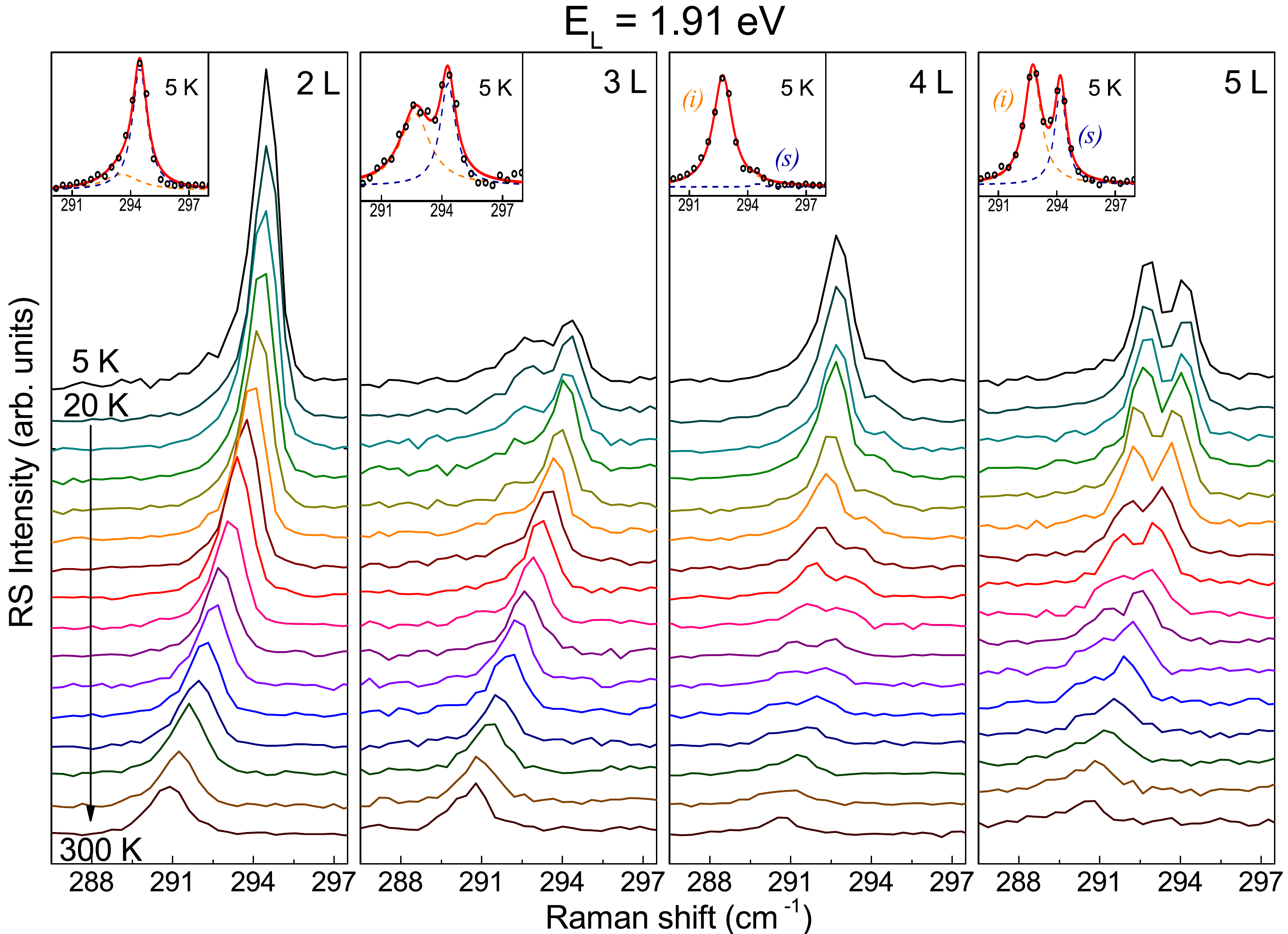}
		\caption{Temperature evolution of high-energy part of the RS spectra of  2 -- 5~L \ce{MoTe2} with the out-of-plane A$_{1\text{g}}$/A'$_{1}$ modes corresponding to the bulk inactive B$^{1}_{2g}$ mode. Laser excitation energy equals 1.91~eV ($\lambda$~=~647.1~nm).}
	\label{fig:647temp}
\end{figure}

A focal point of this paper is the feature apparent at $\sim$291~cm$^{-1}$ in the RS spectra of flakes composed of at least two \ce{MoTe2} layers. Because of symmetries identifing this mode it is labelled as A$_{1\text{g}}$/A'$_{1}$ for few--layer thick structures. The Mulliken notation is exactly the same as the peaks described in the preceding paragraph at $\sim$174~cm$^{-1}$. Therefore, in order to avoid any confusion in further part of this paper, unless otherwise stated, the A$_{1\text{g}}$/A'$_{1}$ label will always refer in what follows to the feature at $\sim$291~cm$^{-1}$. It corresponds to vibrations in which the tellurium atoms in each layer oscillate out-of-phase with respect to the molybdenum atoms (see Fig. \ref{fig:mody}). There are two such modes in bulk \ce{MoTe2}. The atoms in neighboring layers vibrate out-of-phase in the optically inactive B$^{1}_{2g}$ mode while in the infrared-active (A$_{2\text{u}}$) mode the in-phase vibrations occur. As a result none of the modes contribute to the RS in bulk \ce{MoTe2}. Raman-inactive (IR-active) is also the corresponding A''$_{2}$ mode in monolayer \ce{MoTe2}. Additional symmetry elements, like the inversion symmetry center for even or the mirror plane for odd number of layers activate the phonon mode in the material consisting of at least two \ce{MoTe2} layers. Two out-of-plane modes in 2~L \ce{MoTe2} are similar to those present in the bulk. However, due to the symmetry, the A$_{1\text{g}}$ mode, in which the atoms in both layers move out-of-phase with respect to each other becomes Raman-active (see Fig. \ref{fig:mody}). The other mode, A$_{2\text{u}}$, in which the vibrations in both layers occur in-phase, is IR-active. Similar reasoning may be applied to any other $N$-layer structure in which \textit{N} is an even number. The number of Raman-active A$_{1\text{g}}$ modes and the number of IR-active A$_{2\text{u}}$ modes in such cases equals $\frac{N}{2}$ \cite{unified}. If \textit{N} has an odd value, the Raman-active modes must hold the mirror symmetry with respect to the plane crossing Mo atoms in the central layer (see Fig. \ref{fig:mody}). The number of these A'$_{1}$ modes is equal to $\frac{N-1}{2}$ and the number of corresponding IR-active A''$_{2}$ modes amounts to $\frac{N+1}{2}$ \cite{froehlicher}. The above considerations lead in particular to a conclusion that there are: one Raman-active A'$_{1}$ mode and two A''$_{2}$ IR-active modes in 3~L \ce{MoTe2}, two Raman-active A$_{1\text{g}}$ modes and two A$_{2\text{u}}$ IR-active modes in 4~L \ce{MoTe2}, and two Raman-active A'$_{1}$ modes and three IR-active A''$_{2}$ modes in 5~L \ce{MoTe2}. Corresponding normal mode displacements are schematically presented in Fig. \ref{fig:mody}. Two different modes of Raman-active vibrations in 4~L and 5~L samples are referred to as the surface mode (s), and the inner mode (i) \cite{froehlicher, terrones2014new}. The atomic displacements in the corresponding normal modes occur in the outside layers (s) and in the central layers (i), respectively. Based on theoretical calculations \cite{froehlicher} the energy splitting between the modes has been estimated to be $\sim$1.2~cm$^{-1}$. Previous works attributed the splitting mainly to surface effects. The Davydov splitting, which is due to van der Waals interactions between neighbouring layers (being of the order of $\sim$0.1~cm$^{-1}$) is negligible in this case. Experimental study carried out using 1.96~eV excitation show that the inner mode appears in the RS spectra of $N\geq 4$~L \ce{MoTe2} as a faint shoulder on the peak related to the surface mode \cite{froehlicher}. Our results confirm that observation. However, the lower-energy peak related to the bulk inactive B$^{1}_{2\text{g}}$ mode can be observed in the spectra measured at room temperature just as a weak feature (see Fig. \ref{fig:633temp}). In order to improve its visibility by changing the excitation conditions, we have studied the RS as a function of temperature. It is well known that the semiconductor energy band structure changes with temperature with the bandgap usually widening when the temperature is lowered. Temperature modulation therefore allows to form variable resonance conditions without changing the laser excitation energy \cite{livneh16, golasa2017}. The effect of temperature on the investigated RS can be appreciated in Figs \ref{fig:633temp} and \ref{fig:647temp} showing a series of RS spectra excited with 1.96~eV and 1.91~eV light and measured as a function of temperature ranging from 5~K to 300~K on 2 - 5~L \ce{MoTe2}. 

It can be clearly seen in Figs \ref{fig:633temp} and  \ref{fig:647temp} that the lineshape of the investigated spectral features substantially changes with decreasing temperature. Let us start the analysis of the results with the most staggering effects observed for 4~L and 5~L \ce{MoTe2}. For both excitation energies, the lower-energy component of the doublet significantly gains the intensity with decreasing temperature. The effect can be appreciated for both \ce{MoTe2} structures, however, at $T=5$~K it is stronger for 4~L than for 5~L flakes. It is also stronger for 1.91~eV than for 1.96~eV excitation. At  $T=5$~K the lower-energy component of the A$_{1\text{g}}$/A'$_{1}$ modes dominate the high-energy part of the RS spectrum excited with 1.91~eV. Following a theoretical disscusion presented in Ref. \citenum{froehlicher} we attribute the higher-energy component of the investigated peak in 4~L and 5~L to the surface mode (s) of the out-of-plane A$_{1\text{g}}$ or A'$_{1}$ vibrations, respectively, which involve atomic dispacements in the outer \ce{MoTe2} layers (see Fig. \ref{fig:mody}). The lower-energy component is the Raman-active A$_{1\text{g}}$/A'$_{1}$ inner mode (i) in 4~L/5~L \ce{MoTe2}. The energy splitting between the components matches the value predicted within a linear chain model. This points at surface effects as the main factors, which affect the splitting between the surface and the inner mode in the sample. An observation, which goes beyond the previously reported data, concerns the full width at half maximum (FWHM) of both components (see Table \ref{tab:FWHM}).


The lower-energy components of the investigated doublets are broader than their higher-energy counterparts. This may suggest that the former features correspond to more than one vibrational mode. Additional vibrations which may contribute to the lower energy peak are most likely IR active modes. They are expected to be quasi--resonant with the inner modes. Although they should not appear in the RS spectrum, it is known that resonant excitation and/or disorder in the structure can make them Raman-active. This conclusion is supported by the RS from 2~L and 3~L \ce{MoTe2} (see Figs \ref{fig:633temp}, \ref{fig:647temp}). In both cases there is just one Raman-active mode of the out-of-plane vibrations, which is actually the surface mode. However, at low temperature, the lower-energy peak related to the out-of-plane vibrations becomes apparent in the spectrum. Its intensity is lower than the intensity of its higher-energy counterpart but its presence in the RS spectra remains beyond any doubts. As in the case of 4~L and 5~L \ce{MoTe2}, also for 2~L and 3~L \ce{MoTe2} the peaks can be seen more distinctively for 1.91~eV excitation.


\begin{table}[h]
\caption{Full width at half maximum (FWHM) of individual contributions to the A$_{1\text{g}}$/A'$_{1}$ features in the RS spectra collected at $T=5$~K, obtained by fitting the experimental data with a superposition of two Lorentzian peaks.}
        
\centering
    
\begin{tabular}{|c|c|c|c|c|c|c|c|c|}
    
\hline
       Excitation energy & \multicolumn{8}{|c|}{FWHM} \\
       (eV) & 
\multicolumn{8}{|c|}{(cm$^{-1}$)} \\
       
\cline{2-9}
       & \multicolumn{2}{|c|}{ 2 L} & \multicolumn{2}{|c|}{3 L} & \multicolumn{2}{|c|}{ 4 L} & \multicolumn{2}{|c|}{ 5 L} \\   
\cline{2-9}
       & A$_{2\text{u}}$&A$_{1\text{g}}$ & A''$_{2}$&A'$_{1}$ & A$_{1\text{g}}$(i)&A$_{1\text{g}}$(s) & A'$_{1}$(i)&A'$_{1}$(s) \\
       
\hline
        
        1.96  & 3.2 & 2.2 &  2.6 & 0.9  & 1.8 &	1.2 & 1.9  & 1.3 
\\
        1.91  & 2.2 & 0.9 &  1.6 & 0.9  & 1.2  &      0.5  & 1.0 & 0.7
\\
\hline
    
\end{tabular}

\label{tab:FWHM}
\end{table}

In order to analyse the effect of temperature on the RS spectra of few-layer \ce{MoTe2} one can follow the intensities of individual spectral features as a function of temperature. For transparency reasons it is convenient to relate the intensity of the out-of-plane modes to the intensity of the in-plane E$_{\text{g}}$/E' mode, as the in-plane mode is not expected to exhibit any resonant behaviour. Such an approach accounts for small variations of the excitation laser intensity \cite{livneh16}.

\begin{figure}[h]
	\centering  
		\includegraphics[width=0.8\linewidth]{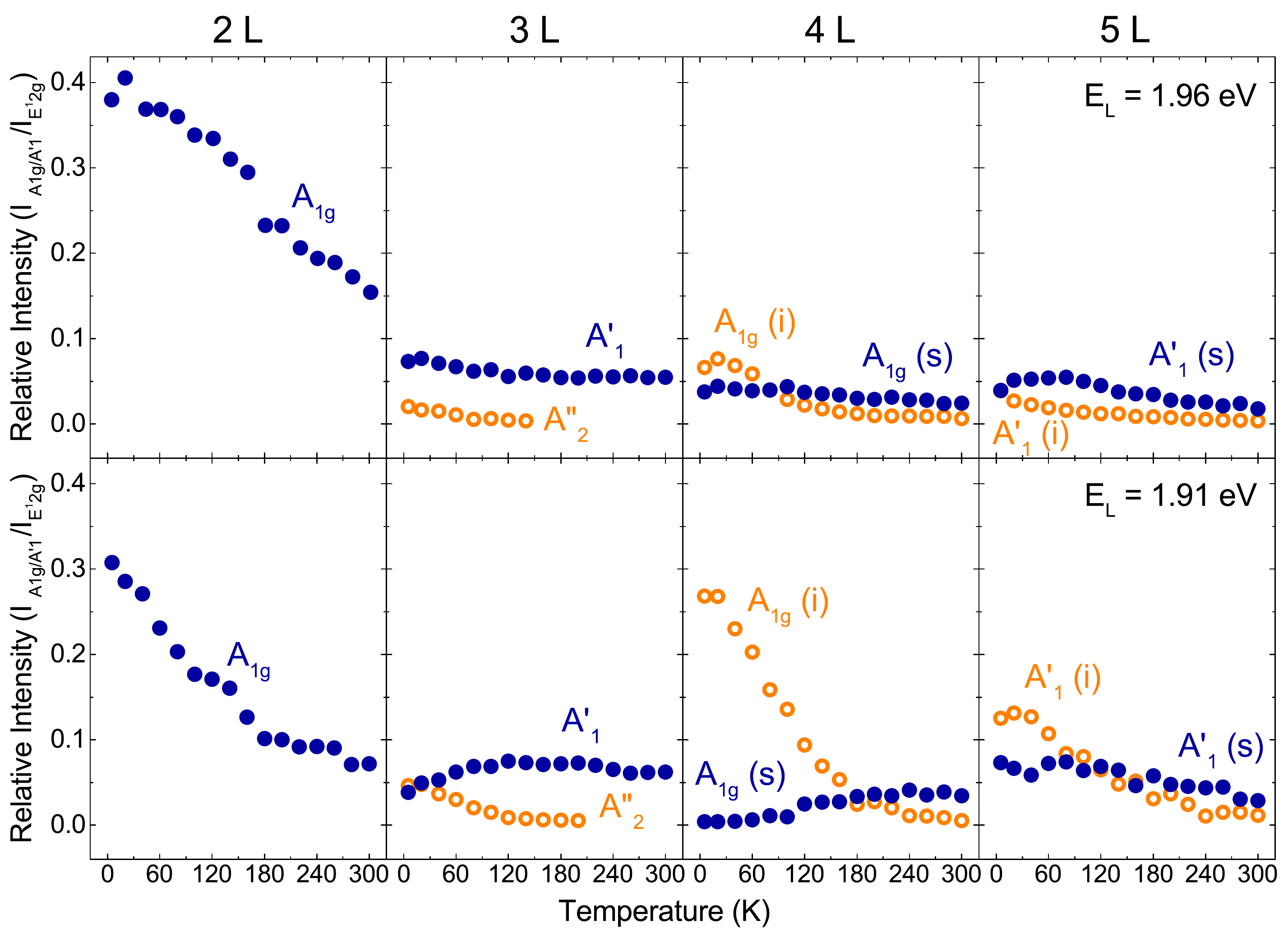}
		\caption{Temperature dependence of the relative intensities 
of the A$_{1\text{g}}$/ A'$_{1}$ - related peaks in the RS spectra of 2 -- 5~L \ce{MoTe2} excited with 1.96~eV ($\lambda$~=~632.8~nm) - top panel and 1.91~eV ($\lambda$~=~647.1~nm) laser light energy - bottom panel.}
	\label{fig:int}
\end{figure}


The temperature dependences of the relative intensities of the A$_{1\text{g}}$/ A'$_{1}$ related peaks observed in the RS spectra of 2 - 5~L \ce{MoTe2} samples under study are presented in Fig. \ref{fig:int}. As previously stated, the most pronunced is the effect of temperature on the RS spectrum excited with 1.91~eV light. The peak due to the A$_{1\text{g}}$(s) mode in 4~L \ce{MoTe2} looses its intensity with decreasing temperature. This behaviour is opposite to the temperature evolution of the peak related to the A$_{1\text{g}}$(i) mode (with a possible contribution from the IR-active out-of-plane mode). This peak is almost seven times more intense at $T=5$~K than at room temperature. A similar effect of temperature can be noticed for spectral features of 5~L \ce{MoTe2}. In this case, the A'$_{1}$(i) mode is three times stronger at $T=5$~K than at room temperature. We interpret the observed effect of temperature on the out-of-plane vibrational modes in terms of the transformation of the \ce{MoTe2} band structure induced by changing the temperature, which we discuss in more detail in the following section.  

\section*{Discussion}


It has previously been noted that the resonance of the exciting light with the electronic transitions in a semiconductor structure can result in strong enhancement of particular RS modes \cite{molas2017raman}. This effect was observed in several semiconductor TMDs including bulk \ce{MoS2} \cite{livneh16}. It was shown that modification of the \ce{MoS2} band structure caused by either the temperature or pressure can lead to significant increase of the intensity of the RS peak due to the out-of-plane Raman-active A$_{1\text{g}}$ mode. We propose a similar explanation for the observed enhancement of the A$_{1\text{g}}$(i) mode in 4~L \ce{MoTe2} reported in this work.

In our case, however, a two-dimensional space of tuning parameters which affect the \ce{MoTe2} band structure is spanned by temperature (\textit{T}) and the number of layers in the sample (\textit{N}). Symbolically, it can be expressed by a formula $E_{i}$(\textit{\textbf{k},T,N}), where \textit{E} stands for the energy, \textit{i} is the band index, and \textit{\textbf{k}} indicates the wave vector. Following this formalism we can say that the results shown in the bottom panel of Fig. \ref{fig:int} clearly demonstrate that the excitation energy of 1.91~eV employed in our experiment corresponds to such a point in the $E_{i}$(\textit{\textbf{k},T,N}) space, where in the vicinity of a particular \textit{\textbf{k}}, $T = 5$~K and $N = 4$ a substantial density of electronic states exists in the \ce{MoTe2} bands linked by optically-active transitions. As also shown in Fig. \ref{fig:int}, tuning \textit{E, N}, or \textit{T} by as little as 50~meV, plus/minus one layer, and 100~K, respectively, while keeping the other two parameters constant, can efficiently quench the intensity of the A$_{1\text{g}}$--mode-related peak in the RS spectrum of few-layer \ce{MoTe2}. It shows how sensitive the resonant RS is as a tool to probe the local density of electronic states in a semiconductor system.

We suggest that the strong enhancement of the A$_{1\text{g}}$(i) mode's intensity in the RS spectrum of 4~L \ce{MoTe2} is associated with electronic transitions taking place at the so-called \textit{M} point of the Brillouin zone of the hexagonal crystal lattice. Indeed, as demonstrated by transmission measurements performed on bulk \cite{wilson1969transition,beal1972transmission} and few-layer 2H--\ce{MoTe2} \cite{song2016,lezama2015indirect}, in the vicinity of 1.91~eV there exists a substantial density of electronic states which gives rise to a doublet of excitonic transitions known as A' and B'. Their labelling intentionally resembles that of the optical transitions at the \textit{K} points of the Brillouin zone as in both cases the local minima and maxima of the conduction and valence band, respectively, are split into pairs of subbands by spin-orbit interaction. Essentially independent of the
sample thickness, the A' transition occurs at room temperature at about 1.73 eV whereas the B' is observed at about 1.98 eV. With decreasing temperature, due to thermal shrinkage of the crystal lattice and resulting effective repulsion of the energy bands, the energies of all excitonic transitions experience a blueshift. For the lowest-energy \textit{K}-point transition between the spin-orbit-split subbands in the conduction and valence bands of 1~L \ce{MoTe2} it amounts to about 0.1~eV as infrared from the comparison of optical data recorded at $4.5$~K \cite{lezama2015indirect} and room temperature \cite{ruppert2014optical}. Worth noting here is that the blueshift of fundamental band-gap transitions is about 0.1~eV when going from room to liquid helium temperature is characteristic of all monolayer TMDs, as shown by temperature-dependent reflectance contrast measurements performed on \ce{WSe2} \cite{arora2015excitonic}, \ce{MoSe2} \cite{arora2015exciton}, and \ce{WS2} \cite{molas2017optical}. To the first approximation, the same temperature evolution as for the \textit{K} points can be assumed also for transitions taking place at other points of the Brillouin zone, as the uniform change in the equilibrium distance between atoms in the \ce{MoTe2} crystal lattice should have a homogeneous impact on the whole band structure. We should then expect the A' and B' transitions to occur at $4.5$~K at about 1.83~eV and 2.08~eV. Although it is difficult to precisely estimate the broadening of these transitions based on the experimental results, presented in Ref. \citenum{song2016,lezama2015indirect} and \citenum{wilson1969transition,beal1972transmission}, it must amount to at least a few tens of meV. It means that with 1.91~eV excitation we most probably probed the high-energy shoulder of the maximum in the local density of electronic states which gives rise to the A' transition. This conjecture also explains the sensitivity of our results to the excitation energy since changes in the resonance conditions should be much more dynamic for the slope than for the close-to-maximum part of the peak in the local density of states. In this respect it would be desirable to extend our study to excitation energies falling in the range from about 1.75~eV to 1.91~eV.


A question arises why only the inner modes are in resonance with 1.91~eV laser excitation, while the surface modes react to it relatively weakly. Definitely a more formal theoretical approach is necessary to account for this observation. We suppose, however, that it may be related to the effect of particular electronic excitation on the polarizability of bonds involved in different vibrational modes. A similar effect was previously studied for Davydov-split modes of the out-of-plane A'$_{1}$(i) vibrations in 3~L \ce{MoTe2} \cite{miranda2017quantum}.
 

\section*{Conclusions}

In conclusion, we have investigated Raman scattering from the out-of-plane vibrational modes (A$_{1\text{g}}$/A'$_{1}$) in few-layer \ce{MoTe2} of thickness ranging from one to five layers. Our temperature-dependent measurements performed with the use of several excitation energies clearly demonstrate a doublet structure of one of the features corresponding to these modes in the RS spectrum at about 291~cm$^{-1}$, especially pronounced in 4~L- and 5~L-thick samples excited with 1.91~eV and 1.96~eV laser light. The lower-- and higher--energy components of the doublet have been identified as the inner and surface modes of the out-of-plane oscillations, respectively. We have showen that for 1.91~eV excitation a strong enhancement of the inner mode's contribution to the RS spectrum is observed for 4~L and 5~L \ce{MoTe2}. Based on temperature evolution of this effect, we qualitatively interpret it as being associated with the resonance between the A$_{1\text{g}}$(i) mode and the A' excitonic transition at the \textit{M} point of the \ce{MoTe2} Brillouin zone, which at $4.5$~K is expected to occur at about 1.83~eV. Taking into account a finite broadening of the A' transition (at the level of few tens of meV), the enhancement of the A$_{1\text{g}}$(i)--mode-related feature's intensity in the RS spectrum we have demonstrated for 1.91~eV excitation can be understood as originating from the onset of resonant conditions (from the high-energy side) reaching their maximum at slightly lower energies, not yet available in the experimental set-up exploited for the measurements reported in this paper. A striking observation about the resonance mentioned above is its extreme sensitivity to the sample thickness which cannot result from the \ce{MoTe2} band structure, which at the \textit{M} point of the Brillouin zone barely changes with the number of layers in the sample. It may hint at important details we have not considered in our simplified interpretation, like, for instance, the thickness and temperature dependence of electron-phonon interactions. Definitely, a much more strict and formal analysis of our experimental data is necessary to fully account for all the results reported in this manuscript, not only qualitatively but also at the quantitative level. We do hope that the present work will stimulate a theoretical study which will shed more light on many aspects of resonant Raman scattering in TMD systems which have not yet been thoroughly understood.


\section*{Methods}

\ce{MoTe2} samples were prepared by an all-dry PDMS-based exfoliation technique \cite{castellanos2014deterministic} of a bulk crystal purchased from HQ graphene and deposited onto a Si/(90 nm) \ce{SiO2} substrate. 
The thickness of studied MoTe$_{2}$ flakes was determined by optical contrast and Raman spectroscopy. 

The unpolarized RS measurements were carried out in the backscattering geometry. 
Raman spectra were taken at various laser excitation wavelenghts: 785~nm (1.58~eV), 625.4~nm (1.98~eV), 632.8~nm (1.96~eV), 647.1~nm (1.91~eV), 514.5~nm (2.41~eV), 532~nm (2.33~eV). The Stokes scattering was measured on 2 -- 5~L \ce{MoTe2} in a temperature range from 5~K to 300~K. The investigated samples were placed on a cold finger of a continuous flow cryostat, mounted on the x--y motorized positioners. The excitation light was focused by means of a 50x magnification long working distance objective. The spot diameter of the focused beam was $\sim$1~$\mu$m. The excitation power focused on the sample was kept at 100 $\mu$W during all measurments to avoid local heating. The RS signal was collected via the same microscope objective, sent through a 0.75 m monochromator and detected with a liquid-nitrogen cooled CCD camera.

\section*{Acknowledgements}

The work has been supported by the European Research Council (MOMB project no. 320590), the EC Graphene Flagship project (no. 604391), the National Science Center, Poland (grant no. 2017/27/B/ST3/00205, 2017/27/N/ST3/01612), the Nanofab facility of the Institut N\'eel, CNRS UGA, and the ATOMOPTO project (TEAM programme of the Foundation for Polish Science co-financed by the EU within the ERDFund).

\section*{Author contributions}

M.G. carried out most optical experiments and preliminary analysed the data. 
K.G., M.R.M., and M. Z. supported the experiments. 
K.N. fabricated the samples under study and performed their characterization.  
M.P. and A. W. contributed to data analysis. 
A.B. supervised the project and performed the final data analysis. 
M.G., K. N. and M.R.M. prepared the manuscript. 
All authors reviewed the manuscript.

\bibliography{bib}
\end{document}